\documentclass{mn2e}
\usepackage{graphicx}
\title[AGES Isolated Galaxies]{The Arecibo Galaxy Environment Survey IX: The Isolated Galaxy Sample}
\author[R. F. Minchin et al.]{R.~F.~Minchin,$^1$\thanks{email: rminchin@naic.edu} R.~Auld,$^2$\thanks{Now at EDF Energy} J.~I.~Davies,$^2$
I.~D.~Karachentsev,$^3$ O.~C.~Keenan,$^2$\newauthor E.~Momjian,$^4$ R.~Rodriguez,$^5$\thanks{Now at University of Texas at San Antonio} T.~Taber$^6$\thanks{Now at Mount Washington Observatory} and R.~Taylor$^7$\\
$^1$ Arecibo Observatory, HC3 Box 53995, Arecibo, PR 00612, UNITED STATES\\
$^2$ Cardiff University, School of Physics and Astronomy, Queen's Buildings, The Parade, Cardiff, CF24 3AA\\
$^3$ Special Astrophysical Observatory, Russian Academy of Sciences,
 Nizhnii Arkhyz, 369167, Zelencukskaya, Karachai-Cherkessia, \\RUSSIA\\
$^4$ NRAO, Dominici Science Operations Center, PO Box O, 1003 Lopezv
ille Rd, Socorro, NM 87801, UNITED STATES\\
$^5$ University of Puerto Rico Humacao, Department of Physics and Electronics, Estaci\'{o}n Postal CUH, 100 Carr. 908, Humacao, \\PR 00791, UNITED STATES \\
$^6$ Vassar College, Physics and Astronomy Department, Ploughkeepsie, NY 12604, UNTED STATES\\
$^7$ Astronomical Institute of the Czech Academy of Sciences, Bo\u{c}ni II 1401/1a, CZ-141 31 Praha 4, Prague, CZECH REPUBLIC}
\begin{document}
\maketitle
\begin{abstract}
We have used the Arecibo L-band Feed Array to map three regions, each of 5 square degrees, around the isolated galaxies NGC 1156, UGC 2082, and NGC 5523. In the vicinity of these galaxies we have detected two dwarf companions: one near UGC 2082, previously discovered by ALFALFA, and one near NGC 1156, discovered by this project and reported in an earlier paper. This is significantly fewer than the $15.4^{+1.7}_{-1.5}$  that would be expected from the field H{\sc i} mass function from ALFALFA or the $8.9 \pm 1.2$ expected if the H{\sc i} mass function from the Local Group applied in these regions. The number of dwarf companions detected is, however, consistent with a flat or declining H{\sc i} mass function as seen by a previous, shallower, H{\sc i} search for companions to isolated galaxies. We attribute this difference in H{\sc i} mass functions to the different environments in which they are measured. This agrees with the general observation that lower ratios of dwarf to giant galaxies are found in lower density environments.
\end{abstract}
\section{Introduction}
The Arecibo Galaxy Environment Survey (AGES) is targetting a wide range of environments from clusters such as Virgo (Taylor et al. 2012; 2013) and Abell 1367 (Cortese et al. 2008) to the Local Void. Isolated galaxies are the least-dense environments targetted that are known, a priori, to contain a single large galaxy and thus a massive dark matter halo.

Isolated galaxies are found in low density environments, outside of groups and clusters. They have long been the object of study, with Karachentseva (1973) defining the original Catalogue of Isolated Galaxies using the 2D criterion of there being no significant companion within 20 optical diameters, where a significant
companion is definied as being one having an optical diameter within a factor of four of the galaxy being studied. More recently, Karachentsev et al. (2011) have taken advantage of our modern knowledge of the 3D distribution of galaxies to construct the Catalogue of Local Orphan Galaxies using 3D isolation criteria. Note that the definition of isolated galaxies is based on the distance to the nearest significant companion, by contrast void galaxies are found inside voids that are much larger than 20 optical diameters but may lie close to a significant companion (particularly along `void filaments', e.g. Alpaslan et al. 2014) and so may not meet the criteria for isolation.

\begin{table*}
\caption{Parameters of the fields searched. ``Distance'' is the adopted distance from the literature to the isolated galaxy, using redshift-independent distance indicators from Karachentsev et al. (2006) for NGC 1156 and from Tully et al. (2009) for UGC 2082 and NGC 5523. ``Upper limit'' is the upper mass limit to which we expect to see companion galaxies, if any were seen above this the galaxy would no longer meet the criteria for isolation. ``Lower limit'' is the lower mass limit to which we expect to see companion galaxies, set by the distance and the RMS noise in the field. ``\% 300 kpc volume sampled'' is the percentage of a 300 kpc radius around the isolated galaxy observed by AGES.}\label{parameters}
\begin{tabular}{lccccccccc}
Field&Parent&Parent&Distance&Physical&RMS&Parent&Upper&Lower&\% 300 kpc\\
&flux&velocity&&area&noise&H\,{\sc i} mass&limit&limit&volume\\
&(Jy km\,s$^{-1}$)&(km\,s$^{-1}$)&(Mpc)&(kpc $\times$ kpc)&(mJy)&(log $M_\odot$)&(log $M_\odot$)&(log $M_\odot$)&sampled\\
\hline
NGC 1156&$75.6$&$379$&7.8&$340\times 270$&0.77&9.03&7.83&6.30&46\\
UGC 2082&$43.5$&$712$&14.7&$640\times 510$&0.81&9.35&8.15&6.87&97\\
NGC 5523&$54.7$&$1050$&20.6&$900\times 720$&0.81&9.74&8.54&7.17&100\\
\end{tabular}
\end{table*}

\begin{figure*}
\includegraphics[width=0.9\columnwidth]{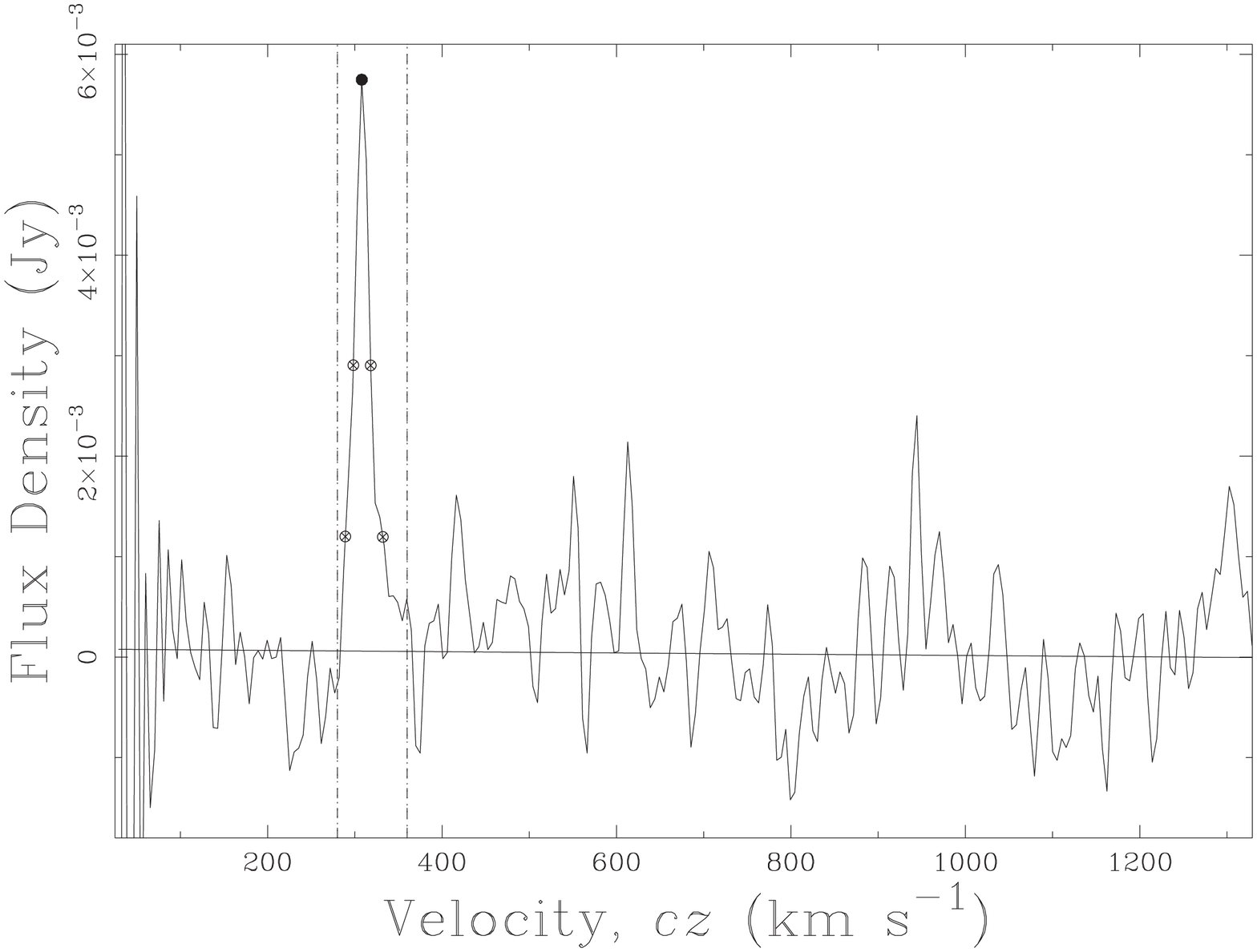}
\includegraphics[width=0.9\columnwidth]{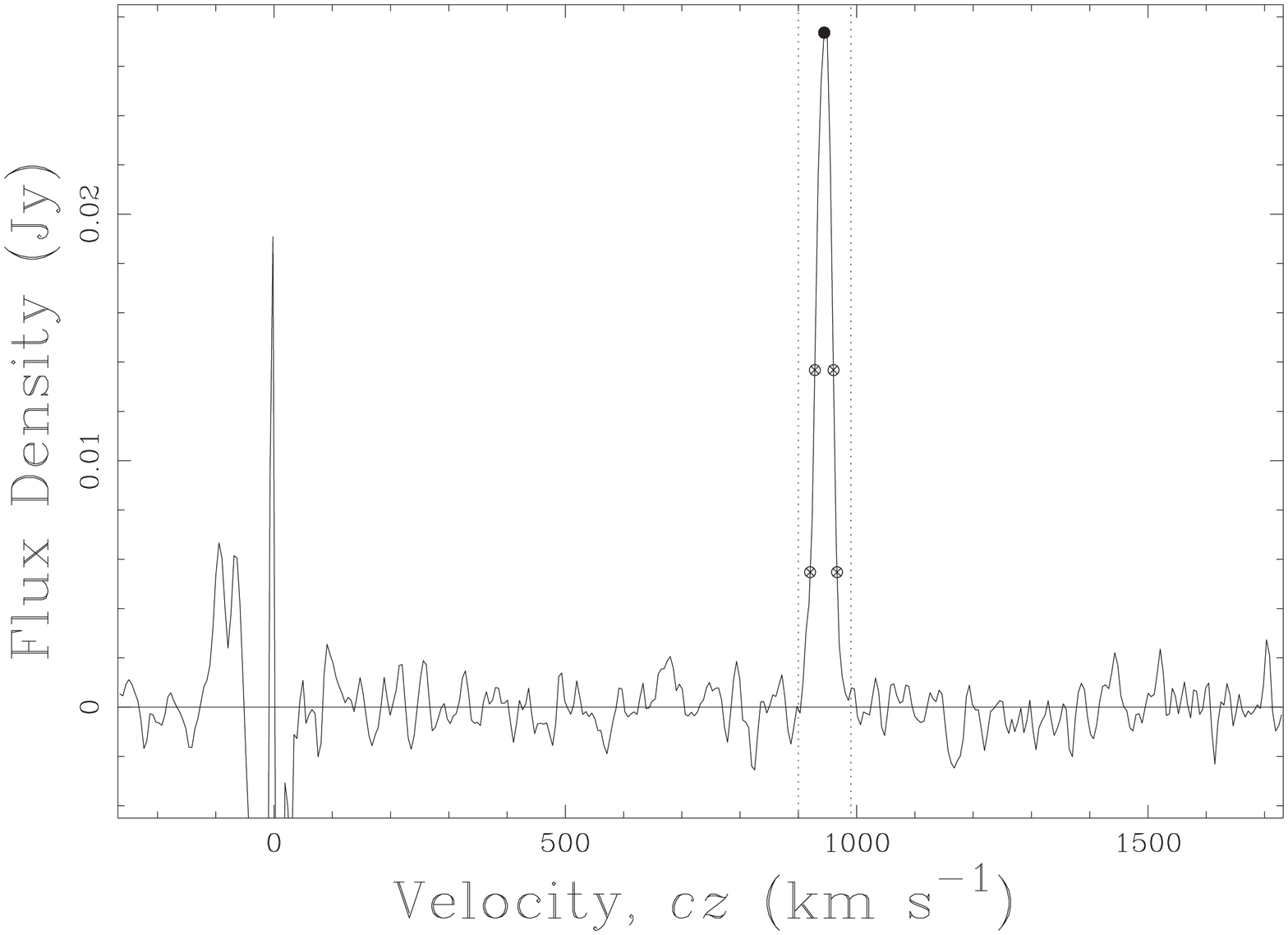}
\caption{AGES spectra of NGC 1156A (left) and UGC 2082A (right). The spectrum for UGC 2082A covers $\pm 1000$ km\,s$^{-1}$ around the velocity of the UGC 2082, while the spectrum for NGC 1156A is truncated at 25 km\,s$^{-1}$ due to problems with data in the Milky Way region.}
\label{spectra}
\end{figure*}

Simulations show that galaxies that appear isolated almost always inhabit the local parent dark matter halo (i.e. they are actually isolated, rather than occupying subhalos of a large, distant parent halo) and that most have been completely isolated since $z = 1$ (Hirschmann et al. 2013). This makes them objects of interest, and isolated galaxies have been the target of recent SDSS studies by Guo et al. (2011) and Wang \& White (2012), as well as the ongoing AMIGA project (Verdes-Montenegro et al. 2005). 

In this paper, we are looking at a sample of three isolated galaxies drawn from Karachentseva (1973)'s catalogue. The three galaxies selected (NGC 1156, UGC 2082, NGC 5523) are all nearby (d $< 21$ Mpc), have large H\,{\sc i} masses ($M_{HI} > 10^9 M_\odot$), and are observable with Arecibo and were observed with AGES. The closeness, combined with the mass of the parent galaxy, gives us a good mass range over which to search for companions. All of our galaxies are also in Karachentsev et al. (2011)'s catalogue, which was not available at the time the sample was defined

One of the major objectives of AGES is to compare the gas-rich galaxy population in different environments. Galaxies in clusters are known to be deficient in gas compared with their counterparts in the field (Giovanelli \& Haynes 1985) due to direct environmental effects such as tidal interactions or ram-pressure stripping. Significant gas deficiency is not seen outside of clusters, so differences in the H{\sc i} mass function (HIMF) between the field and around isolated galaxies are likely to reflect intrinsic differences in the galaxy population. In this paper, we use a technique similar to the dwarf-giant ratio (Roberts et al. 2004, 2007) to compare the number of dwarf companions found in our sample to the numbers predicted from the field HIMF (from ALFALFA; Martin et al. 2010) and the Local Group HIMF (McConnachie 2012). The field HIMF is determined by ALFALFA from $\alpha.40$ catalogue (Haynes et al. 2011), covering a wide range of galactic environments including the Virgo cluster; they find that the inclusion or exclusion of the Virgo cluster gives the same result within the errors.

Observations of a subsample of Karachentseva's galaxies at Arecibo (Haynes \& Giovanelli 1980, 1983, 1984; Hewitt, Haynes \& Giovanelli. 1983) have shown that the H{\sc i} mass depends more on optical size than on morphology. We can use this to determine the region of parameter space in which we do not expect to see any companion galaxies due to the selection criteria. The factor of four in diameter in Karachentseva's selection criteria for isolation is equivalent to 1.2 dex in H{\sc i} mass. As this was used for the selection of our sample, we do not expect to see companions with H{\sc i} masses within 1.2 dex of their parent -- the presence of such sources would have led to the exclusion of their parent galaxies from the sample.

\subsection{Earlier studies}

The only earlier neutral hydrogen survey of the environment of isolated galaxies was by Pisano \& Wilcots (1999), Pisano, Wilcots \& Liu (2002), and Pisano \& Wilcots (2003). This looked at a sample drawn from the Nearby Galaxies Catalogue (Tully 1988) with a low average local bright galaxy density, rejecting those with a peculiar morphology. They also looked at a fairly narrow distance range (22 Mpc $<$ D $<$ 43 Mpc) in order to ensure the parent galaxy was resolved. Of their 41 galaxies, only three are in the catalogue of Karachentseva (1973) and four in the catalogue of Karachentsev et al. (2011), with only one galaxy common to both. The 10 galaxies found to have companions include two of the three in the Karachentseva (1973) catalogue (none are in the Karachentsev et al. 2011 catalogue, although as this was published after the Pisano et al. study it could be that large companions discovered by Pisano et al. led to their exclusion), it thus appears the Pisano et al. criteria are different, rather than less stringent, than the Karachentseva criteria. Our observations extend an order of magnitude deeper on average than the Pisano et al. study.

The AMIGA project (Verdes-Montenegro et al. 2005) also uses a sample drawn from the Karachentseva (1973) catalogue. However, AMIGA applied a low-velocity cutoff that means none of our sample are included in theirs. AMIGA concentrates primarily on the properties of the isolated galaxies themselves, while we are primarily studying their environments. 

The SDSS studies (Guo et al. 2011 and Wang \& White 2012) looked at companions to isolated galaxies, which is the objective of this paper. These studies, however, use a significantly less stringent isolation criterion (differences of 0.5 and 1 magnitude respectively to define signficant companions) and are based on optical data. These differences mean that their results are not directly comparable to ours. Both the SDSS studies found that the number of companions to their isolated galaxies was around a factor of two down from the average number of companions to the Milky Way and M31.

\subsection{Our galaxy sample}

NGC 1156 is an irregular galaxy at a recessional velocity of 379 km\,s$^{-1}$. AGES observations of this galaxy and the surrounding region were reported in Auld (2007) and in Minchin et al. (2010). As in the earlier paper, we adopt the distance estimate of 7.8 Mpc from Karachentsev et al. (2006). Previous H{\sc i} observations (Haynes et al. 1998; Swaters et al. 2002; Springob et al. 2005) give it an H{\sc i} mass (at our adopted distance) of $1.0 \times 10^9 M_\odot$ and $1.1 \times 10^9 M_\odot$ after correction of the flux for source size (based on the optical diameter) and for self-absoprtion. The details of this correction are given in Springob et al. (2005).

UGC 2082 is an Sc galaxy at a recessional velocity of 712 km\,s$^{-1}$. We adopt the distance estimate of 14.7 Mpc from Tully et al. (2009). Previous Arecibo observations reported in Springob et al. (2005) give an H{\sc i} mass (at our adopted distance) of $2.4 \times 10^9 M_\odot$ and $3.1 \times 10^9 M_\odot$ after correction of the flux for source size and self-absorption as for NGC 1156.

NGC 5523 is another Sc galaxy, at a recessional velocity of 1050 km\,s$^{-1}$. We adopt the distance estimate of 20.6 Mpc from Tully et al. (2009). Previous Arecibo observations reported in Springob et al. (2005) give an H{\sc i} mass (at our adopted distance) of $3.6 \times 10^9 M_\odot$ and $4.2 \times 10^9 M_\odot$ after correction for source size and self-absorption as for NGC 1156 and UGC 2082.

\section{Observations and Results}
All three fields were observed with the Arecibo L-band Feed Array (ALFA) in the standard AGES manner, as described by Auld et al. (2007), Cortese et al. (2008) and Minchin et al. (2010). Observations were carried out in 2004-5 (NGC 1156), 2009-10 (UGC 2082) and 2012-13 (NGC 5523). Data for all three fields were taken with the WAPP correlators using the standard AGES setup (Auld et al. 2007; Cortese et al. 2008; Minchin et al. 2010).

The same area ($2.5\degr \times 2\degr$) was observed for all three fields. As they lie at different distances, different physical areas were covered. These areas are given in Table~\ref{parameters}, calculated at the assumed distance of the isolated galaxy. Following Guo et al. (2011) and Wang \& White (2012), we use a fixed radius of 300 kpc from the isolated galaxy and calculate the fraction of the volume within this radius sampled around each galaxy. The differences in distances and in the masses of the parent galaxies also lead to different upper and lower mass limits (assuming a velocity width of 50 km\,s$^{-1}$) for the search in each of the fields. These are presented in Table~\ref{parameters}. 

Fluxes and velocities in Table~\ref{parameters} are measured in a box 3 times the size of the major axis given by the Nasa Extragalactic Database (NED) and corrected based on a 3.4 arcmin Gaussian beam. Note that the formal errors (following Koribalski et al. 2004) are very low due to the brightness of these galaxies; the actual error will be dominated by systematic factors such as baseline fitting, the size of the box used on the map, and the deviation of the Arecibo beam from the Gaussian ideal. We therefore do not quote the formal errors on these quantities to avoid giving the impression of higher accuracy than we have attained. 

The data cubes are searched by eye by two searchers with uncertain detections followed up using the L-band wide receiver at Arecibo Observatory as described below. In order to set a lower limit to our detectable mass, we consider a dwarf galaxy that would be detectable in a by-eye search, i.e. a Gaussian profile with a 5$\sigma$ peak and a FWHM of 50 km\,s$^{-1}$ (higher than either of the galaxies found), truncated at the 20 per cent point (where it reaches the level of the noise). Such an object would have a detection signal to noise ratio (S/N) as defined by Giovenelli et al. (2005) of 5.7. It is likely that fainter sources than this could be detected, particularly at narrower velocity widths, so this sets a conservative limit on our mass range. The noise given in the table is measured over a 1000 km\,s$^{-1}$ range adjacent to the area being inspected. This is $\sim$0.8 mJy in the frequency range of the galaxies, giving lower H{\sc i} mass limits in the range $3.3-3.5 \times 10^4 d^2 M_\odot$. 

\begin{table*}
\caption{Measured parameters of the companion galaxies. ``Flux'' is the total integrated flux of the galaxy. ``Detection S/N'' is calculated following the formula of Giovanelli et al. (2005). ``H\,{\sc i} Mass'' and ``Projected Distance'' are calculated assuming the companion lies at the same distance as the parent galaxy. ``Velocity offset'' is the difference between the parent galaxy's recessional velocity and the companion's recessional velocity.}\label{companions}
\begin{tabular}{lccccccccc}
Galaxy&R.A.&Decl.&Flux&Detection&H\,{\sc i} Mass&$\Delta$V$_{50}$&V&Projected&Velocity\\
&(J2000)&(J2000)&&S/N&&&&distance&offset\\
&&&(Jy km\,s$^{-1}$)&&(log $M_{\odot}$)&(km\,s$^{-1}$)&(km\,s$^{-1}$)&(kpc)&(km\,s$^{-1}$)\\
\hline
NGC 1156A&03:00:39.8&+25:46:56&$0.15\pm 0.04$&9.5&6.33&$20\pm 6$&$308\pm 3$&80&-71\\
UGC 2082A&02:31:21.2&+26:11:58&$0.93\pm 0.05$&44.2&7.68&$32\pm 1$&$944\pm 1$&291&232\\
\end{tabular}
\end{table*}

\subsection{Companions}

Initial analysis of the data cubes was carried out by human inspection of the velocity region around each isolated galaxy. This led to the identification of one candidate near NGC 1156, two candidates near UGC 2082, and one candidate near NGC 5523. One of the candidates near UGC 2082 was detected strongly enough not to require follow-up observing; all of the other candidates were followed up at Arecibo using the single-pixel L-band wide receiver. These follow up observations confirmed the reality of the candidate near NGC 1156, the non-existence of the second (weaker) candidate near UGC 2082, and the non-existence of the candidate near NGC 5523. The measured parameters of these companion galaxies are presented in Table~\ref{companions} and their spectra in Figure~\ref{spectra}.

The companion of UGC 2082, referred to here as UGC 2082A, was discovered by the ALFALFA survey of the anti-Virgo region (Martin et al. 2009), where it is recorded as AGC 122400. It is dubious whether, given its projected distance from UGC 2082 and its separation in radial velocity, it is actually a companion of UGC 2082; we err on the side of including it in our analysis as a companion as this sets a lower limit on the significance of differences between our isolated galaxy sample and HIMFs from other regions (see next section). The companion of NGC 1156, referred to here as NGC 1156A, is a new detection by AGES, previously reported by Minchin et al. (2010) as AGES J030039+254656.

\section{Discussion and Analysis}
We have found two companions to our sample of three isolated galaxies; from Poisson statistics we estimate the error on this measurement to be $\sqrt{2}$, giving us $2 \pm 1.4$ companions. We can compare this to the number we expected from the ALFALFA field HIMF of Martin et al. (2010). We calculate the expected number around each galaxy taking into account that the selection criteria mean we would not expect to see any companion within 1.2 dex of the parent galaxy (upper limit in Table~\ref{parameters}), the lower limit to detectability (lower limit in Table~\ref{parameters}), and the fraction of the virial volume sampled. We make the conservative assumption that companions would be distributed uniformly within the virial volume, if (as is likely) the expected density is centrally peaked this will give us an underestimate of the expected number. The HIMF is normalised to have 3 large galaxies with H{\sc i} masses $> 10^9 M_\odot$.

\begin{figure}
\includegraphics[width=\columnwidth]{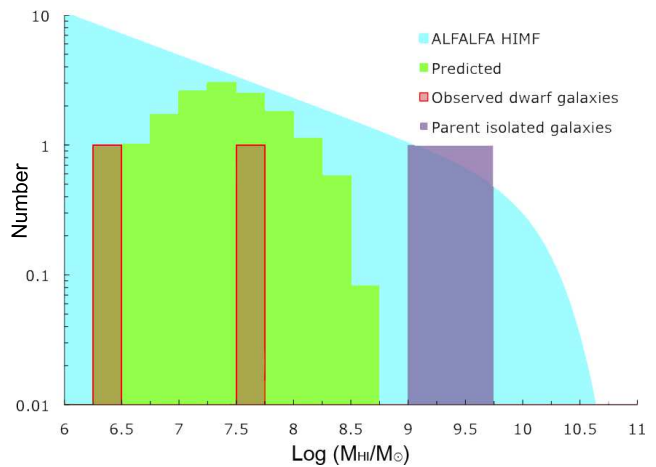}
\caption{Comparison between the AGES Isolated Galaxies mass distribution and the ALFALFA HIMF. The observed H{\sc i} mass distribtion, the ALFALFA HIMF (normalised to the number of large galaxies) and the predicted number of companions based on the ALFALFA HIMF. Bin width for the histogram of number of predicted companions, observed companions and parent galaxies is 0.25 dex; the ALFALFA HIMF is likewise shown in number per 0.25 dex.}
\label{alfalfahimf}
\end{figure}

\begin{figure}
\includegraphics[width=\columnwidth]{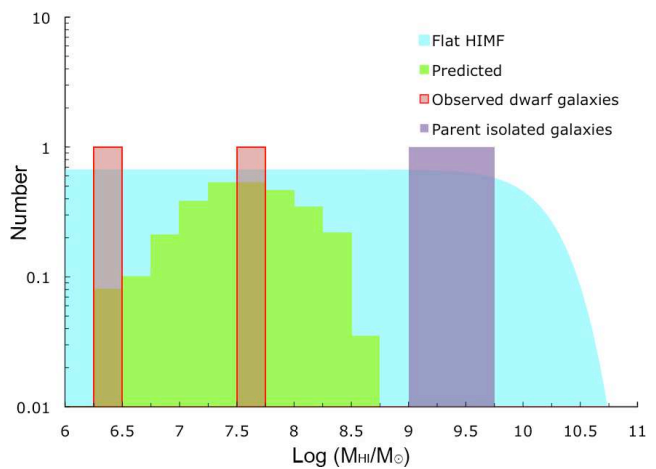}
\caption{Comparison between the AGES Isolated Galaxies mass distribution and a flat HIMF. As Figure~\ref{alfalfahimf} but for a flat HIMF, normalized in the 
same manner to the number of large galaxies.}
\label{flathimf}
\end{figure}


\begin{figure*}
\includegraphics[width=2.0\columnwidth]{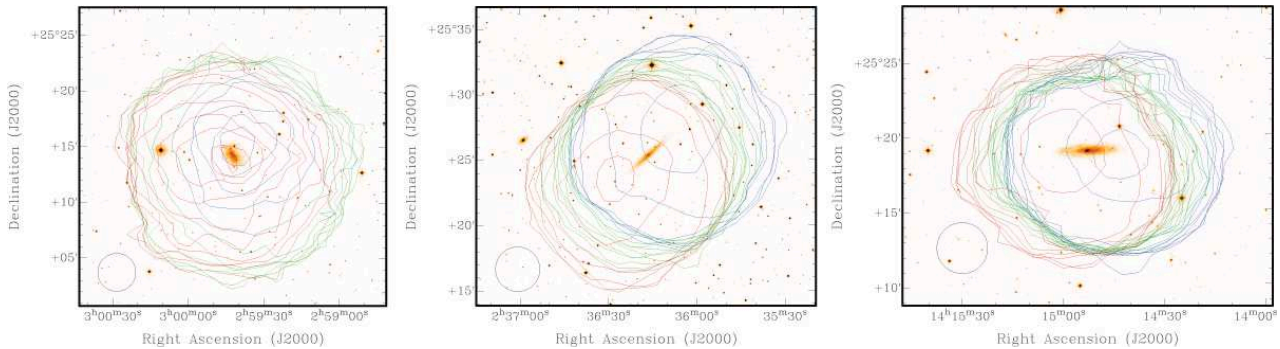}
\caption{Renzograms of the parent isolated galaxies (NGC 1156, left; UGC 2082, centre; and NGC 5523, right) plotted over \textit{B}-band images from the Digitized Sky Survey. Contours are drawn at 10 km\,s$^{-1}$ intervals at the $5\sigma$ level, from low (blue) to high (red) recessional velocity. Beam size is indicated by the black circle to the lower left.}
\label{renzograms}
\end{figure*}

We find (Figure~\ref{alfalfahimf}) that the expected number of companions from the field HIMF with $\alpha = -1.33 \pm 0.02$ is $15.4_{-1.5}^{+1.7}$ at the $1\sigma$ level and $15.4_{-7.8}^{+16.1}$ at the $7\sigma$ level. Adding this in quadrature with the counting error on our observations, we find that we are inconsistent with the field HIMF at greater than $7\sigma$.

 This analysis assumes a uniform mass sensitivity across the range studied, which is not strictly accurate as potential companions can lie in front or behind the isolated galaxy and still be within the 300 kpc radius. However, the maximum difference this could make to sensitivity is less than 10 per cent. To characterise how robust our result is to errors in estimating the mass limit, we look at the effect of raising it. If we assume we can detect all sources with 7$\sigma$ peaks, instead of the 5$\sigma$ peaks used in the calculation of our mass limit, then the expected number of sources falls to $12.8_{-1.1}^{+1.4}$ (ignoring sources detected below that limit), making our result inconsistent at greater than $6\sigma$. Even if we assume our initial mass limit is off by a factor of two, so we now detect galaxies with 10$\sigma$ peaks, we still expect to see $10.6\pm 1.0$ sources, inconsistent with our result at the $5\sigma$ level.

We also compare our sources with a flat HIMF having the same $M^\star$ as the ALFALFA field HIMF and normalised in the same manner to the number of large galaxies, but with $\alpha = -1$ instead of $-1.33$ (Figure~\ref{flathimf}). Here we find that we would expect 2.9 companions, which is consistent with the number found. By altering the HIMF slope to match our observed number, we find that slopes of $\alpha = -1.11$ or steeper are inconsistent at the 95 per cent ($2\sigma$) level.

We further compare our results to the ratio of dwarfs to giants in the Local Group. Using data from McConnachie (2012) for the smaller galaxies, from Kalberla \& Kerp (2009) for the Milky Way, from Braun et al. (2009) for M31, and AGES data from Keenan et al. (in prep.) for M33, we construct the H{\sc i} mass distribution of the Local Group. Normalising this as with the ALFALFA HIMF to have 3 large ($M_{HI} > 10^9 M_\odot$) rather than the 5 found in the Local Group and applying our selection criteria of minimum and maximum mass for companions around each isolated galaxy, we find that we would expect $8.9 \pm 1.2$ dwarf companions in our sample. This differs by more than $3\sigma$ from the observed number. However, if we normalise by the volume of the Local Group ($0.96 \pm 0.30$ Mpc; Karachentsev \& Karachentseva 2006), we would expect $1.4^{+2.8}_{-0.8}$ galaxies. This shows the space density of the dwarfs around the isolated galaxies is similar (after correction for the isolation criteria) to that of dwarfs in the Local Group. This seems, at first sight, to be in tension with the result based on the ratio of companions to large galaxies; however, as we are only considering the volume of the 300 kpc radius sphere around the target galaxy, the space density of large galaxies within this volume is eight times higher than in the Local Group due to their being the target of the survey.

Pisano \& Wilcots (2003) found that, when the parent galaxies were included, their HIMF was declining towards lower masses but flat (over the limited range sampled) when only companions were examined. Our observations are consistent with any declining HIMF in terms of numbers. In terms of the distribution of H{\sc i} masses our small number of sources gives little discriminatory power: the Kolmogorov-Smirnov test shows that we would be inconsistent at the 95 per cent level only with an unfeasibly steeply declining slope having $\alpha > -0.07$. Our observations are therefore consistent with the distribution seen by Pisano \& Wilcots.

The most likely explanation for our finding fewer dwarfs than predicted from the field and local group samples is that there are intrinsically fewer H{\sc i}-rich dwarf companions around isolated galaxies than there are dwarf galaxies per large galaxy in richer environments. This is consistent with the general observation that there is a lower ratio of dwarf to giant galaxies in less dense environments (e.g. Roberts et al. 2007; Tully et al. 2002) and with the interpretation of Pisano \& Wilcots (2003).

\subsection{Evidence for interactions}
We use the H{\sc i} data on the isolated galaxies to look for evidence of interactions, either with the detected companions or with other, undetected galaxies. In Figure~\ref{renzograms} we plot renzograms (named for their inventor, Renzo Sancisi, see Rupen 1999) of all three isolated galaxies. These show contours at the 5$\sigma$ level for each channel of the data cube, with the channels colour-coded to indicate their velocity (with green near the systemic velocity, red at higher recessional velocities, and blue at lower recessional velocities). As can be seen from Figure~\ref{renzograms}, there are no significant disturbances, and smooth rotation is visible as expected in UGC 2082 and NGC 5523 (NGC 1156 is close to face on). There is therefore no evidence from our H{\sc i} data that these isolated galaxies have undergone significant interactions in the recent past.

We also looked at whether the detected companions are likely to have caused any significant disruption to their parent galaxies. We estimate the ratio ($Q$) of the tidal forces ($f_{tidal}$) exerted by the companions on their parent galaxies to the binding forces ($f_{binding}$) of the parent galaxies using the equation from Verley et al. (2007), following the formalism of Dahari (1984):
\begin{equation}
Q = \frac{f_{tidal}}{f_{binding}} = \left(\frac{M_c}{M_p}\right)\left(\frac{D_p}{S}\right)^3
\end{equation}

Where $M_c$ is the mass of the companion galaxy, $M_p$ is the mass of the parent galaxy, $D_p$ is the diameter of the parent galaxy and $S$ is the separation between the parent and companion galaxies. Note that each fraction is dimensionless, so as long as the same mass units are used for the mass of the companion and the parent and the same distance units are used for the diameter of the parent and the separation the specific units used are unimportant.

We assume that the ratio of the total masses is the same as the ratio of the H{\sc i} masses and use optical major diameters for the parent galaxies from NED as the diameter. For the tidal forces to be significant, we would expect $Q \sim 10^{-2}$ or greater; we find that $Q = 4\times 10^{-6}$ for NGC 1156A and $1\times 10^{-5}$ for UGC 2082A. For either of these to be significant would require the H{\sc i} mass to total mass ratio of the companions to be 100 -- 1000 times more than that of the parents. It is therefore not surprising that we do not see evidence for interactions in the renzograms.

\section{Conclusions}
We have shown that the field HIMF is inconsistent at the 7$\sigma$ level with the observed number of companions in the AGES Isolated Galaxies Sample. The number of companions is, however, consistent with a flat HIMF. We also find that the number of companions is inconsistent at the 3$\sigma$ level with the ratio of dwarfs to giants seen in the Local Group. We attribute this inconsistency to a genuine difference between the population of gas-rich dwarf galaxies in the field and around isolated galaxies.

Neither of the companions we have discovered is large enough or close enough to be likely to have a significant tidal effect on its parent, and the H{\sc i} maps of the parent galaxies do not show signs of interactions.

\section*{Acknowledgements}
We thank the anonymous referee for their very helpful comments in the preparation of this paper.

The Arecibo Observatory is operated by SRI International under a cooperative agreement with the National Science Foundation (AST-1100968), and in alliance with Ana G. M\'{e}ndez-Universidad Metropolitana and the Universities Space Research Association.
This research has made use of the NASA/IPAC Extragalactic Database (NED), which is operated by the Jet Propulsion Laboratory, California Institute of Technology, under contract with the National Aeronautics and Space Administration.
The National Radio Astronomy Observatory is a facility of the National Science Foundation operated under cooperative agreement by Associated Universities, Inc.
R.T. is supported by the Tycho project LG14013, the project RVO.67985815 and by the Czech Science Foundation project P209/12/1795. 
I.K. acknoweldges support from the grant of the Russian Scientific Foundation 14-12-00965.

\end{document}